\documentclass[12pt]{article}
\usepackage{graphicx,amsmath,amssymb}
\usepackage{color}

\usepackage{multirow}
\usepackage{subfigure}
\usepackage{setspace}

\newlength{\dinwidth}
\newlength{\dinmargin}
\def\be{\begin{equation}}   
\def\ee{\end{equation}}  
\def\bea{\begin{eqnarray}}                      
\def\eea{\end{eqnarray}}
\def\ch1{$\chi(1^+)$}
\def\lapproxeq{\lower .7ex\hbox{$\;\stackrel{\textstyle                                                    
<}{\sim}\;$}}                                                    
\def\gapproxeq{\lower .7ex\hbox{$\;\stackrel{\textstyle                                                    
>}{\sim}\;$}}

\begin{document}

\begin{flushright}                                                    
IPPP/17/89 \\
\today \\                                                    
\end{flushright} 

\vspace*{0.5cm}

\vspace*{0.5cm}
\begin{center}
{\bf\Large Elastic proton-proton scattering at 13 TeV}\\
\vspace*{1cm}

V.A. Khoze$^{a,b}$, A.D. Martin$^{a}$ and M.G. Ryskin$^{a,b}$\\

\vspace*{0.5cm}
$^a$ Institute for Particle Physics Phenomenology, Durham University, 
Durham, DH1 3LE, UK\\ 

$^b$ Petersburg Nuclear Physics Institute, NRC `Kurchatov Institute', 
Gatchina, St.~Petersburg, 188300, Russia \\
\end{center}

\begin{abstract}
The predictions of a model which was tuned in 2013 to describe the elastic and diffractive $pp$- and/or $p\bar p$-data at collider energies up to 7 TeV are compared with the new 13 TeV TOTEM results. The possibility of the presence of an odd-signature Odderon exchange
contribution is  discussed.

\end{abstract}

\section{Introduction}

 Recently the TOTEM collaboration at the LHC has published the results of the
first measurements at
 $\sqrt{s}$=13 TeV  of the  $pp$ total cross section $\sigma_{\rm tot}=110.6\pm 3.4$ mb 
\cite{TOT1}
and of the ratio of the real-to-imaginary parts of the forward $pp$-amplitude\footnote{
The value $\rho=0.10\pm 0.01$ is obtained from data in the interval $|t|<0.15$ GeV$^2$. If data are used in a more restricted interval $|t|<0.07$ GeV$^2$ (corresponding to the $|t|$ range of the UA4/2 data \cite{ua4}) then $\rho=0.09\pm 0.01$ \cite{TOT2}.}, 
$\rho=$Re$A$/Im$A$=0.10$\pm 0.01$~\cite{TOT2}. 
Here we investigate whether the predictions of a QCD-based multi-channel eikonal model \cite{KMR1,KMR2} are consistent with these measurements.  The measured value of $\rho$ is of particular interest.

Note that the observed value of $\rho$ is quite a bit smaller than that predicted by the
conventional COMPETE parametrization ($\rho=0.13 - 0.14$)~\cite{COMPET,COMPET1}.  The smaller value of $\rho$ may 
indicate either a slower increase of the total cross section at  higher energies or
a possible contribution of the odd-signature amplitude. (Recall that within the COMPETE
parametrization the odd-signature term is described only by secondary Reggeons which dies out
with energy.) Indeed, a $C$-odd amplitude, the so-called Odderon, which depends weakly on energy,
is expected in perturbative  QCD~\footnote{QCD is the $SU(N=3)$ gauge theory which contains the spin=1 particle (gluon) and (for $N>2$) the symmetric colour tensor, $d^{abc}$. Due to these facts in perturbative QCD there {\em exists} a colourless $C$-odd $t$-channel state (formed from three gluons) with intercept, $\alpha_{\rm Odd}$, close to 1.}, 
see in particular \cite{Kwiecinski:1980wb,Bartels:1980pe,Bartels:1999yt}
and for reviews e.g. \cite{Braun:1998fs,Ewerz:2003xi}.
However the naive estimates show that its contribution is rather small;  
say, $\Delta\rho_{\rm Odd}\sim 1\mbox{mb}/\sigma_{\rm tot}\lapproxeq 0.01$~\cite{Rys,LR} at the LHC
energies.

Recall that the Oddeoron was first introduced in 1973~\cite{Lukaszuk:1973nt},
and since then it has been the subject of intensive theoretical
discussion, in particular within the context of QCD. Indeed, there have been several 
attempts to prove its existence experimentally
(see, for example,  
\cite{Braun:1998fs,Ewerz:2003xi,Block} for comprehensive reviews and references).
While the discovery of the long-awaited, but experimentally elusive, Odderon would be 
very welcome news for the theoretical community,
one of our aims here is to evaluate whether the new TOTEM data indicate the presence of Odderon exchange or whether they are consistent with a pure even-signature approach.
 
 To accomplish  this, we compare the new TOTEM results with the predictions of the latest development of our even-signature model~\cite{KMR1,KMR2}.  The model culminated in 2013, and was found to give a successful description of the energy and $t$ behaviour of the total and elastic, $d\sigma_{\rm el}/dt$, proton-proton (proton-antiproton) cross sections, as well as of the diffractive dissociation measured earlier at CERN-ISR, S$p\bar p$S, Tevatron and LHC colliders up to 7 TeV.  The last subset of experimental information is important since, in order to make the analysis more realistic and self consistent, we must include, not only data for the elastic process, but the whole set of soft phenomena, including the diffractive dissociation of the incoming protons; that is, the single and double dissociation processes $pp\to X+p$ and $pp\to X+Y$ where the + sign denotes the presence of a large rapidity gap.

\section{The description of the model}
Let us recall the main features of our `global' approach. To describe the elastic and diffractive data we use a two-channel eikonal model written in the framework of the Good-Walker~\cite{GW} formalism. The QCD-induced Pomeron pole is `renormalized' by enhanced (semi-enhanced)
screening diagrams. The parameters of the 'renormalized' Pomeron, its intercept, $\alpha_P(0)=1+\Delta$, and its effective trajectory slope, 
$\alpha'_P$, were tuned to describe the elastic and diffractive data. We found $\Delta=0.12$ and $\alpha'_P=0.05$ GeV$^{-2}$. The  form factors of the Good-Walker (G-W) eigenstates were correspondingly tuned as well. 

The novel feature of the latest development of the model~\cite{KMR1,KMR2} is that we account for the fact that, due to screening effects, the size, $1/k_P$, of the effective Pomeron decreases with the collider energy. This reflects the growth of the so-called saturation momentum $Q_s^2$ with decreasing $x$.  As a consequence, the couplings, $\gamma_i$, of the G-W eigenstates, $i=1,2$, to the Pomeron depend on the collider energy. At relatively low energy the value of $\gamma_i$ is driven by the size of the particular eigenstate, while at  higher energies it depends mainly on the Pomeron size -- the small size Pomeron interacts with each 
valence quark individually. To reproduce this effect we use simple parametrization 
\begin{equation}
\label{1}
\gamma_i\propto\frac 1{k^2_P+1/r^2_i}\ ,
\end{equation}  
 where $r_i$ is the radius of the state $i$ and
 \begin{equation}
 \label{2}
 k^2_P=k^2_0 ~s^{0.28}\ .
 \end{equation}
Here $\sqrt{s}$ is the $pp$ cms energy.

In this model we see that as $s\to\infty$ both couplings tend to a common value $\gamma_i\to 1/k^2_P$. Thus the probability of low-mass diffractive dissociation decreases with increasing collider energy.  (Recall that, in the G-W formalism the cross section, $\sigma^D_{{\rm low} M}$, for low-mass diffraction\footnote{The cross section for high-mass diffraction is controlled by the triple Pomeron coupling.} $pp\to p+X$,  is proportional to the dispersion of the couplings $\gamma_i$.) This allows the model to reproduce the unexpectedly low value of 
$\sigma^D_{{\rm low} M}=2.6\pm 2.2$ mb for $M<3.4$ GeV observed by TOTEM at $\sqrt s=7$ TeV~\cite{TOT-D}. Indeed, in our model we find 3.8 mb.

Recall that  at CERN-ISR energies it was observed that the ratio 
$\sigma^D_{{\rm low} M}/\sigma_{\rm el}\simeq 0.3$, while at 7 TeV
it becomes about 0.1. This behaviour of the ratio with increasing collider energy was not able to be reproduced by earlier models.

\begin{figure}
\begin{center}
\vspace{-8cm}
\includegraphics[width=20cm]{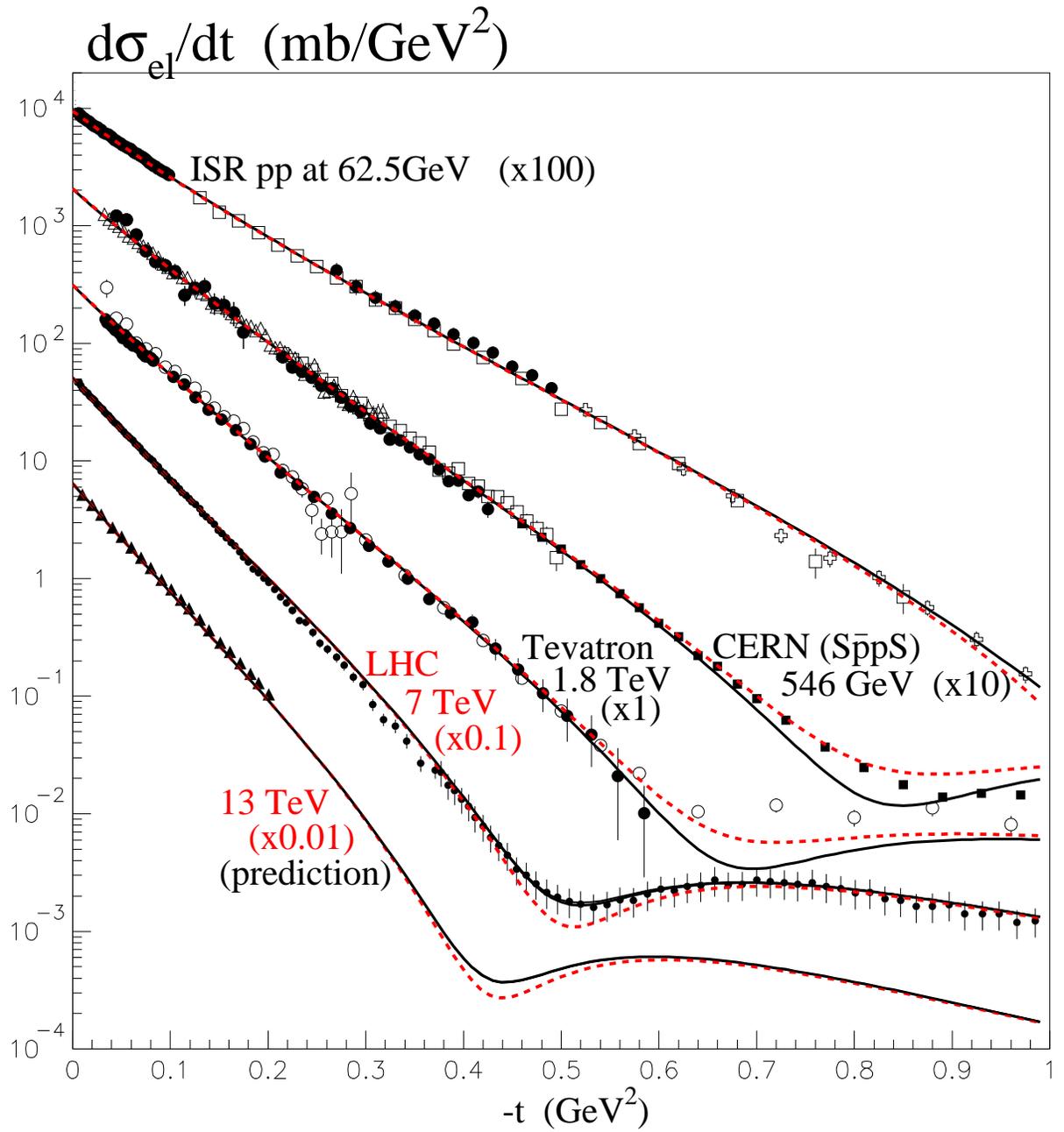}
\end{center}
\caption{\sf The dependence of the $pp$ (or $p\bar p$) elastic cross section on the momentum transferred square $t$ compared with the present data (see \cite{KMR2} for the references), and the prediction for $\sqrt s=13$ TeV.  The continuous curves correspond to the original model \cite{KMR1,KMR2}, whereas the dashed curves show the effect of including an Odderon contribution as described in the text. The 13 TeV data are from \cite{TOT2}.} 
\end{figure}

\section{Predictions of the model}
The model has a small number of parameters and is intended to give an overall description of elastic and quasi-elastic (i.e. diffractive) $pp$ high-energy interactions.
With the limited number of parameters, the model is more reliable in the small $|t|$ region (before the dip).  At larger $|t|$, in particular in the dip region and beyond, the predictions are sensitive to small changes in the values of the parameters.   

In Fig.1 we show the description of the elastic proton-(anti)proton differential cross section data, together with the prediction for $\sqrt{s}$=13 TeV using the final (2013) version  of the model~\cite{KMR1} without any additional tuning.
In Table 1 we give the values of the total cross sections, the ratio $\rho=$Re$A$/Im$A$, $\sigma_{\rm el}$ and the $t$-slope, $B_{\rm el}$ at $t=0$ and the effective slope measured in the interval $0.05<|t|<0.15$ GeV$^2$.
The model predictions \cite{KMR1} $\sigma_{\rm tot}=111.2$ mb and $\sigma_{\rm el}=29.5$ mb at 13 TeV should be compared to the observed values of $110.6\pm 3.4$ mb and $31.0\pm 1.7$ mb \cite{TOT1}.

\begin{table}
\begin{center}
\begin{tabular}{|c|c|c|c|c|c|}
\hline
 $\sqrt s$ &  $\rho$ &    $\sigma_{\rm tot}$ & $\sigma_{\rm el}$ &  $B_{\rm el}(0)$ &      $B_{\rm el}(|t|=0.05-0.15\mbox{GeV}^2)$\\
 (TeV) &   & (mb)  &  (mb) &   (GeV$^{-2}$) & (GeV$^{-2}$)  \\
 \hline
   0.546 &    0.128 &    62.5 &     12.8  &   14.7   &  14.9  \\ 
   1.8 &   0.123  &   77.1   &  17.4  &   16.8   &  16.7 \\   
    2.76 &   0.121  &   83.2   &  19.5 &    17.6 &     17.5 \\ 
   7.  &  0.117 &    98.8 &     24.9  &   19.7  &   19.4\\ 
  8.  &  0.116  &  101.3   &  25.8   &  20.1    & 19.7  \\    
{\bf  13.}  & {\bf 0.113} & {\bf 111.2} &{\bf 29.5} & {\bf 21.4} & {\bf 21.0} \\ 
 100. &    0.102 &    166.2  &   51.5 &     29.4 &     29.8 \\ 
\hline
\end{tabular}
\end{center}
\caption{The values of the observables given by the model~\cite{KMR1}.} 
\end{table}

 \subsection{The $t$ dependence of the elastic slope}
 
 Note that the $t$ dependence of the differential cross section $d\sigma_{\rm el}/dt$ cannot be  described by a pure exponent. The behaviour is more complicated.   The proton form factor and the 
pion-loop insertion into the Pomeron trajectory, as well as absorptive corrections, all result in some variations of the `local' $t$-slope.   
The pion loop and the proton form factor lead to the slope decreasing with $|t|$, whereas on the other hand absorptive effects lead to the slope (before the first diffractive dip)  increasing with $|t|$.  A detailed  discussion, and references, of these effects can be found, for example, in~\cite{KMRt}. 

Therefore we have shown in Table 1 not only the slope $B_{\rm el}(0)$ at $t=0$ but also the effective slope measured in the 
$0.05 <-t<0.15$ GeV$^2$ interval. At the LHC energies (7$-$13 TeV) the effective slope from the $0.05- 0.15$ GeV$^2$ interval is a bit smaller than the slope at $t=0$, 
mainly due to the pion loop and the form factor effects. However, at higher energies the effects due to absorptive corrections become more important (in this $t$ interval).  Indeed, we see from Table 1 that the value of the effective slope  (last column in Table 1) exceeds the slope at $t=0$ for 100 TeV.
 
 Note that the slope at 13 TeV is determined from data in the interval $0.01<|t|<0.2$ GeV$^2$. The observed value $20.36\pm 0.19$ GeV$^{-2}$ \cite{TOT1} is therefore best compared to our model prediction of 21.0 GeV$^{-2}$.
 The `discrepancy' is discussed in Section 
\ref{sec:4}, in particular in footnote 11.

\subsection{Real part of the (even-signature) amplitude}
Recall that the model includes  {\em only} even-signature amplitudes. Actually we first calculate just the imaginary part of the amplitude. The real part of elastic amplitude can be obtained using dispersion relations.
However, the model did not include secondary Reggeon contributions. Thus we cannot describe the cross sections at relatively low energies
 which enter the dispersion relation. Therefore we use the following more simplified 
 approach to calculate the real part of the amplitude\footnote{This approach is used not only at $t=0$ but also at $t\neq 0$ to calculate the real part of the amplitude which fills the diffractive dips in the elastic cross sections $d\sigma_{\rm el}/dt$ of Fig.1.}.
 
 The even-signature amplitude
 \begin{equation}
 A^{(+)}~=~(A(s)+A(u))/2~~\propto ~~s^\alpha+(-s)^\alpha\ ,
\end{equation}   
where at high energies the Mandelstam variable $u\simeq -s$.
Thus we obtain
\begin{equation}
\rho~~\equiv~~ \frac{\mbox{Re} A}{\mbox{Im} A}~=~\tan(\pi(\alpha-1)/2)\ .
\label{4}
\end{equation} 
Due to the absorptive corrections (induced in this model by the eikonal) the energy dependence of the amplitude is not equal to that given by single Pomeron exchange. In central collisions (i.e. at small values of the impact parameter $b$) the corrections are stronger. 
Therefore we transform (\ref{4}) to impact parameter space and calculate
the value of $\alpha(b)$ as
 \begin{equation}
 \alpha=\frac{d\ln A(b)}{d\ln s}
 \end{equation}
 at each point of $b$ space. That is, we use the signature factor 
 \be
 \eta=i+\tan(\pi(\alpha-1)/2)
 \ee
  accounting for the `effective' value of intercept $\alpha(b)$ 
  which describes the energy behaviour of the amplitude at fixed value of $b$ and  depends on $b$~\footnote{The $b$ dependence of the imaginary and the real parts of the amplitude were shown in Fig.6 of ~\cite{KMR2}.}. At high energies this approach provides sufficiently good accuracy, better than about 0.003 in $\rho$.
 Indeed, describing the lower energy contribution by the exchange of secondary Reggeons
 (mainly the $f_2$ and $\omega$ trajectories) we see that this term dies out as $1/\sqrt s$. Indeed using the COMPETE parametrization~\cite{COMPET} we find that already at $\sqrt s=541$ GeV this contribution to $\rho=$Re$A/$Im$A$ i
s   less than 0.002.

Returning to the high energy behaviour of the amplitude, we note that COMPETE uses a simplified parametrization motivated by Froissart asymptotics
\be
\frac{1}{s}{\rm Im}A(s,t=0)~~=~~c~{\rm ln}^2(s/s_0)~+~P~+~R(s)
\label{eq:COMP}
\ee
where $c$ and $P$ are constants and $R(s)$ corresponds to the contribution of the secondary Reggeons. However, even at 13 TeV we are far from asymptotics; the coefficient $c=0.272$ mb is much less than that corresponding to  the Froissart limit of $c\simeq 60$ mb. In general, we expect the actual pre-asymptotic energy behaviour to be more complicated than (\ref{eq:COMP}).  In our model \cite{KMR1,KMR2} the {\it asymptotic} behaviour is also of the form $\sigma_{\rm tot}\to c'{\rm ln}^2 s$, but since the couplings to the G-W eigenstates, $\gamma_i$ of (\ref{1}), have their own $s$ dependence, we predict\footnote{We emphasize that the even-signature amplitude generated by our model is an analytic function that satisfies the usual dispersion relation which determines the real part of the amplitude in terms of the energy behaviour of the imaginary part.} a lower value $\rho=0.113$ at 13 TeV, in comparison to $\rho=0.131$ of COMPETE \footnote{The value $\rho=0.131$ corresponds to the parameters presented by the PDG in \cite{COMPET}. However, this set of parameters gives a cross section $\sigma_{\rm tot}=105.6 $mb at 13 TeV, which is too small as compared to the TOTEM value of 110.6 mb. The COMPETE parameters which give $\sigma_{\rm tot}=110.6$ mb yield $\rho=0.135$.}.

The predictions for $\rho$ are shown in Table 1, and by the continuous curve in Fig.~\ref{fig:2}.  Even without an odd-signature contribution, the model could reasonably 
well describe the currently most
  precise experimental results for $\rho=$Re$A$/Im$A$, namely $\rho=0.135\pm 0.015$ at 541 GeV~\cite{ua4} and 
$\rho=0.10\pm 0.01$ at 13 TeV~\cite{TOT2}.
However, as we shall show below, the addition of a small `Odderon' (odd-signature) term would 
certainly improve the description of the data since 
it would enlarge the value of $\rho$ for the $p\bar p$ data at 541 
GeV and reduce the real part of $pp$ amplitude at 13 TeV.

\subsection{Inclusion of the odd-signature Odderon contribution}

Until now we have only accounted for the even-signature contribution to the amplitude. On the other hand, besides the odd-signature terms given by the $\rho,\omega$ Reggeons, there exists in perturbative QCD an odd-signature $t$-channel state (the QCD Odderon) with intercept close to 1 \cite{Kwiecinski:1980wb}~-~\cite{Ewerz:2003xi}. The exchange of such a state will produce an odd-signature amplitude which is almost purely real and which decreases very weakly with increasing energy. The simplest example is 3-gluon exchange. In the Born (i.e., lowest $\alpha_s$) approximation
we may consider the exchange of three gluons between the valence quarks of the colliding protons. It is the presence of the symmetric colour tensor $d_{abc}$ which allows the formation of this $C$-odd signature 3-gluon state. Recall that, as shown in \cite{Kwiecinski:1980wb,Bartels:1980pe}, the real and virtual corrections to this Born amplitude cancel each other to good accuracy. So the lowest $\alpha_s$ approximation is not too bad.

To estimate the effective coupling of such an odderon to a proton in $pp$ scattering a simplified model was used in \cite{Rys}. The corresponding impact factor was calculated assuming that the proton is formed by three valence quarks in an oscillator potential whose parameter is chosen to reproduce the known electromagnetic radius of the proton, see \cite{Rys,LR}. This leads to a pure real odd-signature amplitude \footnote{The normalization is taken to satisfy Im$A(s,t=0)~=~s\sigma_{\rm tot}$.}
\be
\frac{1}{s}{\rm Re}A^{(-)}~
\simeq~0.8~{\rm mb}.   
\label{8}
\ee

In this subsection we study the possible effects of such an amplitude added to our previous predictions. Recall that the elastic amplitude was originally written in impact parameter, $b$, space in the form
\be
A(b)~=~i\left(1-e^{-\Omega (b)/2} \right),
\label{un}
\ee
which is the exact solution of the elastic $s$-channel unitarity equation
\be
2{\rm Im} A(b)~=~|A(b)|^2+G_{\rm inel} (b),
\label{10}
\ee
where $\Omega (b)$ is the opacity of the proton and $G_{\rm inel}$ accounts for the inelastic channels.
The new odd-signature term should be added to $\Omega (b)$ so that $\Omega$ contains an additional imaginary part.

In order not to introduce too many new parameters, the secondary Reggeon contributions were taken with couplings given by the COMPETE parametrization, and $t$ dependence described by the usual dipole form factor $1/(1-t/0.71~{\rm GeV}^2)^2$.  Moreover, the couplings (of the secondary Reggeon terms and the new Odderon term) to the different G-W eigenstates are chosen to be the same. We parameterize the $t$ dependence of the Odderon term by exp$(B_{\rm Odd}t)$ with  the slope for the amplitude $B_{\rm Odd}=6$ GeV$^{-2}$. Using, another value of the slope, or instead of the exponential, a pole or dipole parametrization, gives essentially the same result, except for small changes in the dip region.

\begin{figure}[h]
 \begin{center}
 \includegraphics[trim=0 0cm 0 7.6cm,scale=0.5]{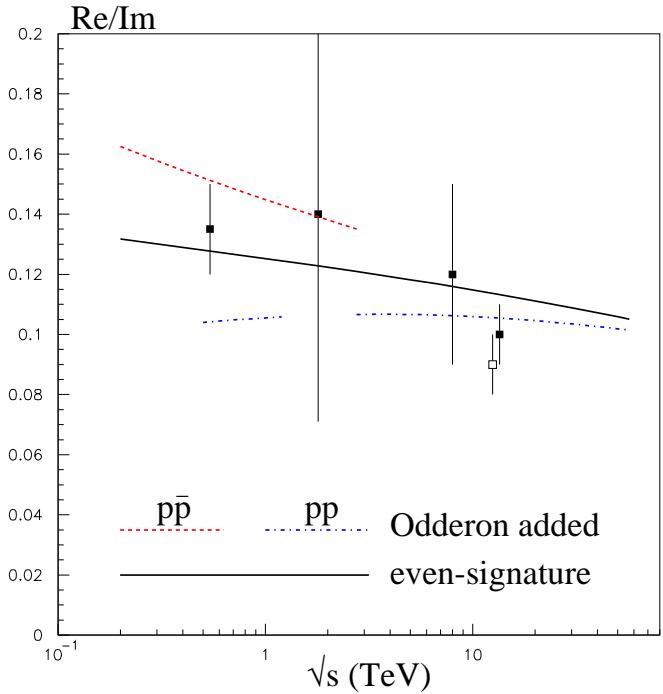}
 \caption{\sf The energy dependence of the $\rho=$Re$A$/Im$A$ ratio. The data are taken 
 from~\cite{ua4,e710,TOT8,TOT2}; the first two data points correspond to $p\bar{p}$ scattering and the last points to $pp$ scattering. At 13 TeV we also show by the open square the value of $\rho$ 
 obtained under the same conditions as that used by the UA4/2 group (see footnote 1).  The values of $\rho$ given by the model~\cite{KMR1} are shown by the solid curve. The dashed curves include a {\em possible} QCD Odderon contribution calculated as described in the text.}
 \label{fig:2}
 \end{center}
 \end{figure}

As expected, the secondary Reggeon contributions are already small at ISR energies and are practically invisible for $\sqrt{s}\gapproxeq$ 500 GeV. The Odderon contribution, with a coupling of 0.8 mb is also quite small. However enlarging the coupling by a factor of two is not excluded by the oversimplified model of \cite{Rys}. In this case we obtain a larger real part in $p{\bar p}$ scattering and a smaller $\rho$ in $pp$ scattering. 
Taking a QCD Odderon coupling of 2.8 mb (in the normalization of eqs.~(\ref{8})$-$(\ref{10})) and the 
slope\footnote{Note that the $C$-odd and isospin=0 state does not couple to the pion. Thus the Odderon only feels the centre of the proton, and not the pion cloud. Therefore it is reasonable to assume that the Odderon slope, $B_{\rm Odd}$, is lower than that for the even-signature (Pomeron) amplitude.}
$B_{\rm Odd}=6$ GeV$^{-2}$, we find
the values of $\rho$ shown by the dashed curves in Fig.~\ref{fig:2}.  For $p\bar{p}$-scattering at $\sqrt{s}$=541 GeV we now have $\rho$~=~0.15, close to the $1\sigma$ experimental limit: $\rho=0.135\pm 0.015$ \cite{ua4}. Simultaneously, the prediction for $pp$ scattering at 13 TeV decreases to $\rho$~=~0.107 in better agreement with the TOTEM measurement \cite{TOT2}.  Note that at the higher energy the Odderon contribution gives a smaller effect due to the stronger screening caused by Re$\Omega (b)$; that is, the second term in (\ref{un}) dies out.

\section{Discussion   \label{sec:4}}
As seen from Table 1 and the accompanying text, within the error bars the model predictions \cite{KMR1}
are in agreement with all the new TOTEM data \cite{TOT1,TOT2}.
Even without an odd-signature contribution the model could reasonably 
well describe the currently most
  precise experimental results for $\rho=$Re$A$/Im$A$, namely $\rho=0.135\pm 0.015$ at 541 GeV~\cite{ua4} and 
$\rho$ comprised between $0.09$ and $0.10(\pm 0.01)$ at 13 TeV~\cite{TOT2}, as shown by the continuous curve in Fig. 2.  Recall that the same model successfully describes \cite{KMRt} the deviation from a pure exponential behaviour of the cross section $d\sigma_{\rm el}/dt$ that was measured precisely by TOTEM at 8 TeV \cite{TOT8}.

However, the addition of a small `QCD Odderon' (odd-signature) term would 
certainly improve the description of the data since 
it would enlarge the value of $\rho$ for the $p\bar p$ data at 541 
GeV and reduce the real part of $pp$ amplitude at 13 TeV.  Fig. 1  shows that the presence of the Odderon (with a reasonable coupling) is invisible in $d\sigma_{\rm el}/dt$ for $|t|\lapproxeq 0.2$ GeV$^2$, but that it is noticeable in the region of the diffractive dip (improving the description of the 546 GeV and 1.8 TeV data).

Recall that in any reasonable model (in particular QCD) the high energy Odderon contribution weakly decreases with energy due to the lower intercept, $\alpha_{\rm Odd}(0)>\alpha_{\rm even}(0)$, and the stronger absorptive corrections which increase with $s$ \cite{FFKT,KMROdd}. The fact that the Odderon contribution to $\rho$ increases as the energy decreases puts an {\it upper} limit on the Odderon amplitude coming from the UA4/2 $p\bar{p}$ 
value of $\rho=0.135\pm 0.015$. As seen from Fig. 2, the parameters we have chosen are already close to the upper limit\footnote{We do not consider here the `maximal Odderon' \cite{Martynov:2017zjz}, since it was shown \cite{KMROdd} that the maximum Odderon amplitude is inconsistent with unitarity.}.

To conclude, we repeat, that even without the odd-signature term, the model of \cite{KMR1,KMR2} predicts the new 13 TeV TOTEM data \cite{TOT1,TOT2} reasonably well.  The largest disagreement is the value of the elastic
slope. The model predicts $B_{\rm el}=21.0$ instead of $20.4\pm 0.2$ GeV$^{-2}$ quoted by TOTEM \cite{TOT1}. \footnote{Note that at both $\sqrt{s}$=2.76 \cite{TOT2.76} and 13 TeV the observed slopes are too close to the values measured earlier at a smaller energy, and which are in good agreement with the model value (see Table 1).
We list below the values of the slopes $B_{\rm el}$ in units of GeV$^{-2}$ at the relevant energies:\\
-- at 1.8 TeV we have $16.98\pm 0.25$ CDF \cite{CDF} or $16.99\pm 0.47$ E710 \cite{e710} (model gives 16.8),\\
-- while at 2.76 TeV we have $17.2\pm 0.3$ TOTEM \cite{TOT2.76} (model gives 17.5);\\
-- at 7 TeV we have $19.73\pm 0.14\pm 0.26$ ALFA-ATLAS \cite{ATLAS7} or $19.9\pm 0.3$ TOTEM \cite{T7} (model gives 19.7),\\
-- while at 13 TeV we have $20.36\pm 0.19$ TOTEM \cite{TOT2} (model predicts 21.0). }

The inclusion of the Odderon does improve the calculated value of $\rho$ at 13 TeV. The Odderon contribution is practically invisible in $d\sigma_{\rm el}/dt$ at low $|t|$ values, but will reveal itself in the region of the diffractive dip where the imaginary part of the even-signature amplitude vanishes. It will be very interesting to study $d\sigma_{\rm el}/dt$ in the dip region and to check the low $|t|$ slope $B_{\rm el}$ in future ALFA-ATLAS and TOTEM experiments. As seen in Fig. 2, the difference between the $\rho$ values for $pp$ and $p\bar{p}$ in the region of $\sqrt{s}$=900 GeV caused by the Odderon can be significant. Precise data in this region would be informative.

 \section*{Acknowledgements}

VAK acknowledges  support from a Royal Society of Edinburgh  Auber award. MGR thanks the IPPP of Durham University for hospitality.

\end{document}